\documentclass[aps,prd,preprint,superscriptaddress,tightenlines,nofootinbib]{revtex4}

\usepackage{graphicx}
\usepackage{dcolumn}
\usepackage{bm}

\begin{document}

\preprint{CLNS 08/2020}       
\preprint{CLEO 08-04}         

\title{Precision Measurement of the Mass of the $\bm{h_c(^1P_1)}$ State of Charmonium}

\author{S.~Dobbs}
\author{Z.~Metreveli}
\author{K.~K.~Seth}
\author{A.~Tomaradze}
\affiliation{Northwestern University, Evanston, Illinois 60208, USA}
\author{J.~Libby}
\author{A.~Powell}
\author{G.~Wilkinson}
\affiliation{University of Oxford, Oxford OX1 3RH, UK}
\author{K.~M.~Ecklund}
\affiliation{State University of New York at Buffalo, Buffalo, New York 14260, USA}
\author{W.~Love}
\author{V.~Savinov}
\affiliation{University of Pittsburgh, Pittsburgh, Pennsylvania 15260, USA}
% ...
\collaboration{CLEO Collaboration}
\noaffiliation

\date{May 28, 2008}

\begin{abstract}

A precision measurement of the mass of the $h_c(^1P_1)$ state of charmonium has been made using a sample of 24.5 million $\psi(2S)$ events produced in $e^+e^-$ annihilation at CESR.  The reaction used was $\psi(2S) \to \pi^{0} h_{c}$, $\pi^{0} \to \gamma \gamma$, $ h_{c} \to \gamma \eta_{c}$, and the reaction products were detected in the CLEO-c detector.
  Data have been analyzed both for the 
inclusive reaction and for the exclusive reactions in 
which $\eta_c$ decays are 
reconstructed in fifteen hadronic decay channels. Consistent results
are obtained in the two analyses. 
The averaged results of the present measurements are $M(h_c)=3525.28\pm 0.19 (\mathrm{stat})\pm 0.12(\mathrm{syst})$ MeV,  and 
$\mathcal{B}(\psi(2S)\to\pi^0h_c)\times\mathcal{B}(h_c\to\gamma\eta_c)$=
(4.19$\pm$0.32$\pm$0.45)$\times 10^{-4}$.
Using the $^3P_J$ centroid mass, $\Delta M_{hf}(1P)\equiv \left<M(\chi_{cJ})\right>-M(h_c)=
+0.02\pm0.19\pm0.13$ MeV.

\end{abstract}

\maketitle

The large body of experimental data for
 the spectroscopy of the charmonium ($c\bar{c})$ states has
provided detailed information about the QCD interactions between a 
quark and
an antiquark. A convenient and transparent realization 
of the interaction is in terms of a potential which is generally assumed to
consist of a Coulombic part attributed to a vector one gluon
exchange, and a less well understood confinement part. 
In analogy with QED,
the spin--dependence of the interaction is attributed to the Breit--Fermi 
reduction of the one--gluon vector exchange, which leads to spin--orbit 
($\mathbf{L}\cdot \mathbf{S}$), 
tensor $(T)$ and spin--spin $(\mathbf{S}_{1}\cdot \mathbf{S}_{2}$) potentials.    
 The confinement part  
is generally assumed to be Lorentz scalar and no spin--spin dependence arises from it. The mass splitting of the triplet
$1P$ charmonium states into $\chi_{c0}(^{3}P_{0})$, $\chi_{c1}(^{3}P_{1})$
and $\chi_{c2}(^{3}P_{2})$ is determined
by the $(\mathbf{L}\cdot \mathbf{S})$ and $(T)$ terms of the potential, and the 
$(\mathbf{S}_{1}\cdot \mathbf{S}_{2}$) term
determines the hyperfine or triplet--singlet splitting. If the $q\bar{q}$ hyperfine interaction receives 
no contribution from the confinement part, and is only due to the Coulombic term
in the potential, it is a contact interaction in the lowest order,
and it is identically zero for all $L\ne0$, i.e.,
$\Delta M_{hf}(1P)\equiv M(^{3}P)-M(^{1}P)=0$. 
The triplet $^{3}P_{J}$ states are well established, and the mass of their
spin--weighted centroid is 
 $\left<M(^{3}P_J)\right>=[M(\chi_{c0})+3M(\chi_{c1})+5M(\chi_{c2})]/9=
3525.30\pm0.04$ MeV \cite{pdg}. 
The singlet state $h_c(^{1}P_{1})$ was not 
 identified until very recently \cite{hcold,fermilab}.
Although the identification of the triplet centroid mass $\left<M(^{3}P_J)\right>$ with the unperturbed triplet mass $M(^3P)$ has been questioned \cite{martinrichards}, it is necessary to make a precision measurement of the mass of $h_c$ irrespective of how $M(^3P)$ is determined.

Two recent experiments have reported 
identification of $h_c$ and measured its mass.
The CLEO measurement \cite{hcold} was made by means of the 
isospin-forbidden reaction 
\begin{equation}
\psi(2S)\to\pi^0 h_c, \pi^0\to\gamma\gamma,~h_c\to\gamma\eta_c
\end{equation}
using 3 million $\psi(2S)$ produced in $e^+e^-$ annihilations.
The $h_c$ was identified as the enhancement in the mass spectrum 
of recoils against $\pi^{0}$. Two different kinds of analysis 
of the data were done. In the \textit{inclusive analyses}
$h_c$ decays were identified by loose constraints on either the energy of
the E1 photon from $h_c$ decay, or the mass of $\eta_c$. 
In the \textit{exclusive analysis} no constraint was placed on $E(\gamma)$.
 Instead,  $\eta_c$ events were
reconstructed in seven different hadronic decay channels of $\eta_c$.
The combined significance level of the $h_c$ observation was $>$6 $\sigma$,
and the quoted mass was $M(h_c)$=3524.4$\pm$0.6$\pm$0.4 MeV.

The Fermilab E835 measurement \cite{fermilab} made scans of antiproton
energy for the reaction, $\bar{p}p\to h_c\to \gamma\eta_c$, 
$\eta_c\to \gamma\gamma$. The results from the year 1997 scan 
and the year 2000 scan were 
combined to obtain $M(h_c)$=3525.8$\pm$0.2$\pm$0.2 MeV.
The significance level of 
$h_c$ observation was  $\sim 3\sigma$. No evidence was found for 
 $h_c$ in the previously reported reaction  
$\bar{p}p\to h_c\to \pi^{0}J/\psi$~\cite{fermilab1}.

If it is assumed that
 $M(^{3}P)$=$\left<M(^{3}P_{J})\right>$, the above two measurements
lead to $\Delta M_{hf}(1P)$=+0.9$\pm$0.6$\pm$0.4 MeV (CLEO), and
$\Delta M_{hf}(1P)$=--0.5$\pm$0.2$\pm$0.2 MeV (FNAL).
While both results are statistically consistent with the prediction
of $\Delta M_{hf}(1P)$=0, it is important to understand any deviation from it, 
and its origin. 

In this Letter we report a much improved measurement of the reaction in Eq.~(1) using nearly an order of magnitude larger sample of $N(\psi(2S))=24.5\pm0.5$~million \cite{npsip} obtained at the Cornell Electron Storage Ring with $e^+e^-$~annihilations at a center of mass energy corresponding to the $\psi(2S)$ mass of 3686~MeV \cite{pdg}.  The CLEO-c detector was used for the detection of the reaction products.

The CLEO-c detector \cite{cleodetector}, which has a cylindrical geometry, 
consists of a CsI electromagnetic calorimeter, an inner vertex drift chamber, 
a central drift chamber, and a ring-imaging Cherenkov (RICH) detector, 
inside a superconducting solenoid magnet with a 1.0 T magnetic field. 
The detector has a total acceptance of 93$\%$ of $4\pi$, 
 photon energy resolutions of 2.2$\%$ at $E_{\gamma}$=1 GeV, and 
5$\%$ at 100 MeV, and charged particle momentum resolution of 0.6$\%$ at 1 GeV.

The event selection criteria common to both the inclusive and exclusive
analyses are the following. The events were required to have at least three 
electromagnetic showers and two charged tracks meeting the standard CLEO 
quality and vertex criteria \cite{dbr}. The acceptance region was defined 
as $|\cos \theta| \le 0.93$, except that recoil $\pi^{0}$ candidates 
were reconstructed
using photons only in the good barrel region, $|\cos \theta| \le 0.81$.
For showers it was required that  $E_\gamma(\mathrm{barrel})>30$ MeV, and
 $E_\gamma\mathrm{(endcaps})>50$ MeV, where the endcap region is defined as $0.85 < |\cos \theta| < 0.93$. 
The events accepted for $\gamma \gamma$ decays of $\pi^0$ and $\eta$
were required to have $M(\gamma\gamma)$ within $\pm$15 MeV 
of $M(\pi^{0}$)=135.0 MeV and $M(\eta)$=547.5 MeV, respectively~\cite{pdg} .
It was further required that there be only one $\pi^0$ in the event with the recoil mass in the expected region of $h_c$ mass, $3526\pm30$~MeV. 
 These candidates were fit kinematically with $M(\gamma\gamma)$ constrained to the $\pi^0$
and $\eta$ masses to improve energy resolution.
To distinguish charged pions, kaons and protons a log-likelihood 
criterion including $dE/dx$ and information from the RICH detector
was used.

In the \textit{inclusive analysis}, in order to remove neutral 
pions from $J/\psi$ decays following 
$\psi(2S)\to\pi^+\pi^-J/\psi$ and $\pi^0\pi^0J/\psi$, events were 
rejected with $\pi^+\pi^-$ recoil mass in the range
$M(J/\psi)=3097\pm15$ MeV and $\pi^0\pi^0$ recoil mass in the range 
$M(J/\psi)=3097\pm40$ MeV. Similarly, events with 
the invariant mass of all charged particles, 
$M(\mathrm{all~charged})>3050$~MeV,
as well as events with recoil mass against $\gamma\gamma$ in the range 
$M(J/\psi)=3097\pm40$ MeV, were rejected to remove decays through 
the $\chi_{J}$ states.

\begin{figure}
\includegraphics*[width=3.3in]{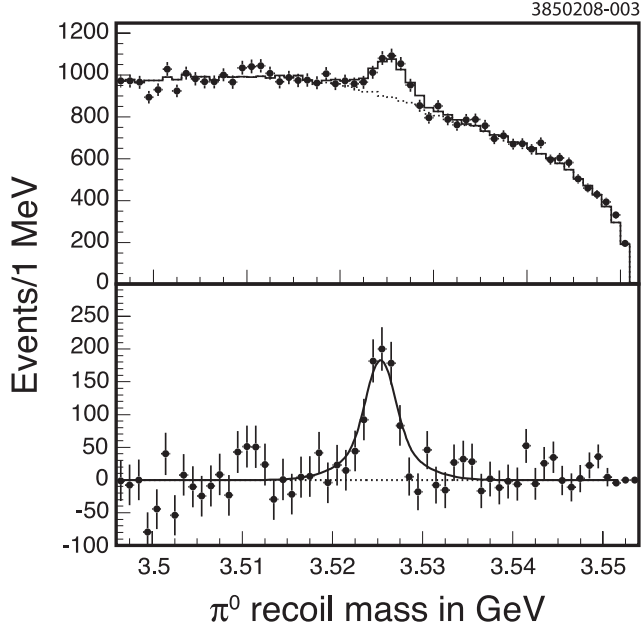}
\caption{Spectra of recoil masses against $\pi^0$ in the inclusive 
analysis: (top) full spectrum, (bottom) background subtracted spectrum.}
\vspace*{-0.3cm}
\end{figure}

For the inclusive analysis it is required
that the energy of the E1 photon in $h_c\to\gamma\eta_c$ be in the
expected range $E(\gamma)=503\pm35$ MeV.
It is also required that there be only one 
such photon in the event. Further, this candidate photon was rejected 
if it made either a $\pi^{0}$ or $\eta$ with any other photon in the event.

\begin{table}

\setlength{\tabcolsep}{5pt}

\caption{Results for the inclusive and exclusive analyses for the reaction $\psi(2S)\to\pi^0h_c\to\pi^0\gamma\eta_c$.  First errors are statistical, and the second errors are systematic. As described in the text,  $\mathcal{B}_1\times\mathcal{B}_2$ for the exclusive analysis is based on $165\pm19$ counts.}

\begin{center}
\begin{tabular}{lcc}
\hline \hline
 & Inclusive & Exclusive \\ \hline
Counts & $1146\pm118$ & $136\pm14$ \\
Significance & $10.0\sigma$  & $13.2\sigma$ \\
$M(h_c)$, MeV &3525.35$\pm$0.23$\pm$0.15  &3525.21$\pm$0.27$\pm$0.14   \\
$\mathcal{B}_1\times\mathcal{B}_2\times10^4$ & $4.22\pm0.44\pm0.52$ & $4.15\pm0.48\pm0.77$ \\
\hline \hline
\end{tabular}
\end{center}
\vspace*{-0.3cm}
\end{table}

The mass spectra of $\pi^{0}$ recoils are shown in Fig.~1,
with the full spectrum in the top panel, and the background subtracted
spectrum in the bottom panel.
When the requirement $E(\gamma)=503\pm35$ MeV is not imposed, a spectrum of the background is obtained with nearly twenty times
larger yield and no apparent $h_c$ enhancement, as is expected because of
the small product branching fraction 
$\mathcal{B}_1(\psi(2S)\to\pi^0h_c)\times\mathcal{B}_2(h_c\to\gamma\eta_c)$
$\approx$4$\times$10$^{-4}$. 
To remove the small $h_c$ contribution in the above background, events were removed if they had a photon with $E(\gamma)=503\pm50$~MeV.
In the fit of the $h_c$ spectrum in
Fig.~1~(top) this background shape was mapped to the full spectrum 
with just one normalization parameter. The peak shape used consists of
a Breit--Wigner function with an assumed width of 0.9 MeV (same as $\Gamma(\chi_{c1})$),
convolved with the instrumental resolution function
 obtained by fitting the
Monte Carlo (MC) simulation of the data.
The $\chi^2/$d.o.f. of the fit is 54/52.
In the MC simulations the angular distribution for the E1 photon
was assumed to be  $(1 + \cos^2 \theta)$.
The overall efficiency determined from the 
MC sample is $\epsilon$=11.1\%. The results of the fit, and
$\mathcal{B}_{1}\times\mathcal{B}_{2}=N(h_c)/(\epsilon\times N(\psi(2S))$)
 are listed in Table I.

In the \textit{exclusive analysis} no constraint on 
$E(\gamma)$ was imposed. Instead, for the decays
$\psi(2S) \to \pi^{0} h_{c}$, $ h_{c} \to \gamma \eta_{c}$, $\eta_c \to X$, 
$\eta_{c}$ candidates were reconstructed in fifteen different
decay modes, $X$, with multiplicities of 2 to 6. These modes were used 
because they had significant yields in the direct decays 
$\psi(2S) \to \gamma \eta_{c}$. Several of them, marked with (*),
were utilized for the first time. 
These decay channels are: $p\bar{p}$, 
$\eta\pi^{+}\pi^{-}$ ($\eta\to \gamma\gamma$),
$\eta\pi^{+}\pi^{-}$ ($\eta\to \pi^{+}\pi^{-}\pi^{0}$), 
$K_SK^{+}\pi^{-}$, $K^{+}K^{-}\pi^{0}$, 
$\pi^{+}\pi^{-}\pi^{+}\pi^{-}$, 
$K^{+}K^{-}\pi^{+}\pi^{-}$, 
$K^{+}K^{-}K^{+}K^{-}$, 
$\pi^{+}\pi^{-}\pi^{+}\pi^{-}\pi^{+}\pi^{-}$,
$K^{+}K^{-}\pi^{+}\pi^{-}\pi^{+}\pi^{-}$,
(*)$\eta K^{+}K^{-}$($\eta\to \gamma\gamma$),
(*)$p\bar{p}\pi^{0}$;
(*)$\pi^{+}\pi^{-}\pi^{0}\pi^{0}$,
(*)$p\bar{p}\pi^{+}\pi^{-}$, 
(*)$\pi^{+}\pi^{-}\pi^{+}\pi^{-}\pi^{0}\pi^{0}$.

The decay chain in Eq.\ (1) as well as the above $\eta_c$ decays were 
identified from the reconstructed charged particles, and 
$\pi^0$'s and $\eta$'s. 
For $\eta$ decays to $\pi^+\pi^-\pi^0$, it was required that the invariant 
mass be within 30~MeV of the nominal mass $M(\eta)$=547.5 MeV \cite{pdg}.
For $K_S^0$ decaying into a $\pi^+\pi^-$ pair, it was required that the invariant mass of 
the pair be within 10 MeV of the nominal 
mass $M(K_S^0)$=497.6 MeV \cite{pdg}, 
and information  about vertex displacement was used to reject 
random $\pi^+\pi^-$ combinations.  
The $\psi(2S)\to\pi^+\pi^-J/\psi$  events were rejected with $\pi^+\pi^-$
recoil mass in the range $M(J/\psi)=3097\pm15$ MeV.

The entire decay sequence was reconstructed for each $\eta_c$ decay 
channel.  A kinematically constrained fit was done for each event to 
take advantage of energy-momentum conservation, and it was required that the 
$\chi^2$ of the 4C fit be less than 15.  
The mass of the $\eta_c$ candidates was required to be  within 
30 MeV of the nominal mass of $M(\eta_c)$=2980 MeV \cite{pdg}.  If multiple $\eta_c$ candidates were found in an event, only the one with the smallest $\chi^2$ was retained.

\begin{figure}
\includegraphics*[width=3.3in]{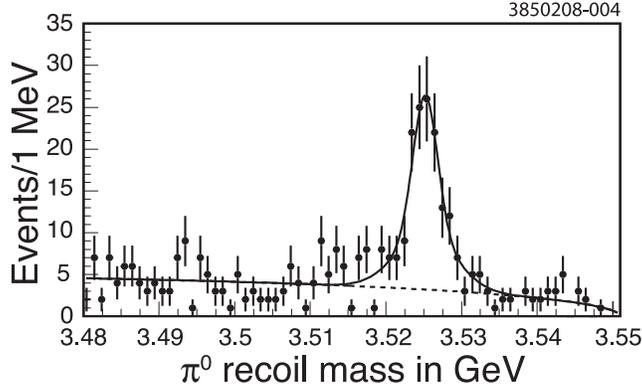}
\caption{Summed distribution of recoil masses against $\pi^0$ in the exclusive 
analysis with 15 decays channels of $\eta_c$. See text for details.}
\vspace*{-0.3cm}
\end{figure}

The $\pi^{0}$ recoil mass distribution for each decay channel was fitted
separately using the instrumental resolution
shape determined from MC simulation, convolved with a Breit--Wigner function
of assumed width $\Gamma(h_c)$=0.9 MeV. 
The ARGUS shape \cite{argus} was used to parameterize the background.
The fitted number of counts from 
individual decays range from 1 to 30.  The summed distribution
was fitted in the same way. The fit is shown in Fig.~2.

The product branching ratio 
$\mathcal{B}_1(\psi(2S)\to\pi^0h_c)\times\mathcal{B}_2(h_c\to\gamma\eta_c)$
is related to the observed counts in the different decay channels $\eta_c\to X$ as the average
\begin{equation}
\left< \frac{N(X,h_c)/\epsilon(X,h_c)}{N(X, \mathrm{direct})/\epsilon(X,\mathrm{direct})} \right>_{X}=
\frac{\mathcal{B}_1\times\mathcal{B}_2}
{\mathcal{B}(\psi(2S) \to \gamma \eta_{c})}
\end{equation}

In order to minimize systematic errors in the evaluation
of Eq.~(2) it is desirable to construct $\eta_c \to X$ decays in the same manner
for $\eta_c$ from $h_c$ and  $\eta_c$ from direct decay of $\psi(2S)$.
We do so by placing a window
of $\pm$7 MeV around $M(h_c)$ in $\pi^{0}$ recoil.
The spectrum for the hadronic system mass for each individual decay
channel was then reconstructed and fitted in the same manner for decays
through $h_c$ as for the direct decays. The fits were done using 
a Breit--Wigner function with $\Gamma (\eta_c)$=26.5 MeV \cite{pdg}, convoluted with 
the experimental resolution function as determined by the MC
simulation for that channel, and parametrized as a double Gaussian.
The background in each case was parametrized using a polynomial.
The number of counts in individual decays ranged from 37$\pm$11 ($p\bar{p}$)
to 1052$\pm$74 ($\pi^{+}\pi^{-}\pi^{+}\pi^{-}\pi^{0}\pi^{0})$,
with a total $\Sigma N(X,\mathrm{direct})$=4043$\pm$127. 
The corresponding total
 $\Sigma N(X,h_c)$ was 165$\pm$19.
This is larger than $\Sigma N(X,h_c)$ obtained by fitting the 
$\pi^{0}$ recoil spectrum for  $M(h_c)$ measurement, and has 
correspondingly larger efficiency.

The efficiencies $\epsilon(X,h_c)$ and $\epsilon(X,\mathrm{direct})$ were determined from
MC simulations separately for each channel.
As expected, it was found that the ratios of efficiencies, $R(X)=\epsilon(X,\mathrm{direct})/\epsilon(X,h_c)$ were essentially independent of $X$, and had the average value $\left<R\right>=2.36\pm0.17$.  This allows us to obtain from Eq.~2.
\begin{equation}
\frac{\mathcal{B}_1\times\mathcal{B}_2}
{\mathcal{B}(\psi(2S) \to \gamma \eta_{c})}
=\frac{\sum N(X,h_c)}{\sum N(X,\mathrm{direct})} \times \left<R\right>
=0.096\pm0.013.
\end{equation}

\begin{figure}
\includegraphics*[width=3.3in]{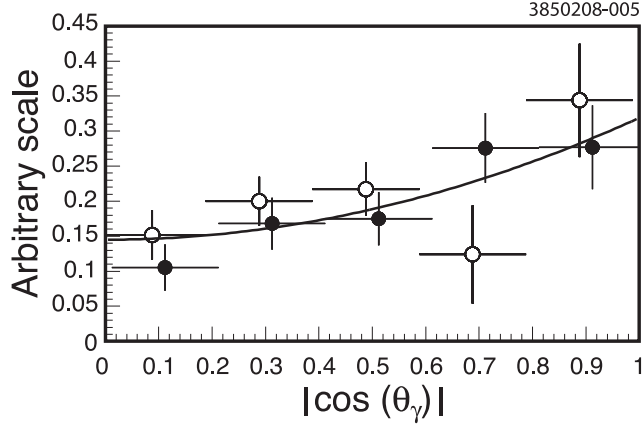}
\caption{Angular distributions of the photons from $h_c \to \gamma \eta_c$.
Circles and solid points denote results from
inclusive and exclusive analyses, respectively.
The curve shows $N(1 + \alpha \cos^2\theta)$ distribution,
corresponding to $\alpha$=1.20, as explained in the text.}

\end{figure}

Using the summed counts above, and the recently measured CLEO value,   
$\mathcal{B}(\psi(2S) \to \gamma \eta_{c})$=(4.32$\pm$0.67)$\times$10$^{-3}$ \cite{hadron2007}, 
we obtain $\mathcal{B}(\psi(2S) \to \pi^{0} h_{c})\times\mathcal{B}(h_{c}\to \gamma \eta_{c})$=
(4.15$\pm$0.48(stat))$\times$10$^{-4}$.

The angular distributions of the E1 photons in both inclusive and
exclusive analyses were
obtained by fitting separately the $h_c$ peak in the data for 
different angular ranges.
The results are shown in Fig.~3.  
The distributions were fitted with the function $N(1 + \alpha \cos^2 \theta)$.
The fits give $\alpha_{incl}$=0.87$\pm$0.65 ($\chi^2$/dof=3.9/3)
 and $\alpha_{excl}$=1.89$\pm$0.94  ($\chi^2$/dof=1.8/3).
In order to take the average of the results from inclusive and exclusive 
analyses, the exclusive events were removed from the inclusive sample.
The average of the values from inclusive and exclusive analyses is 
$\alpha_{average}$=1.20$\pm$0.53, and the curve in Fig.~3 illustrates it.
This is consistent with $\alpha$=1 expected for an E1 transition
from $h_c(J^{PC}=1^{+-})$ to $\eta_c(J^{PC}=0^{-+})$.

\begin{table}
\label{tb:systematic}
\caption{Summary of estimated systematic errors and their sum in quadrature.  
N/A means not applicable.}
\begin{center}
\begin{tabular}{l|cccc}
\hline \hline
 &  \multicolumn{2}{c}{$M(h_c)$, MeV} & \multicolumn{2}{c}{$\mathcal{B}_1\times \mathcal{B}_2\times10^4$}\\
Systematic uncertainty in & Incl. & Excl. & Incl. & Excl. \\
\hline
$N(\psi(2S))$ & N/A & N/A & 0.08 & N/A\\
$\mathcal{B}(\psi(2S)\to\gamma\eta_c)$ & N/A & N/A & N/A & 0.66\\
Background shape & 0.10 & 0.01 & 0.26 & 0.15\\
$\pi^0$ energy calibration & 0.08 & 0.08 & N/A & N/A  \\
$\pi^0$ signal shape & 0.03 & 0.01 & 0.14 & N/A \\
$h_c$ width & 0.03 & 0.02 & 0.27 &  $<0.01$ \\
efficiency  & N/A & N/A & 0.20  &  0.22 \\
Binning, fitting range & 0.03 & 0.03 & 0.08 & 0.27\\
MC input/output & 0.05 & 0.11 & N/A & N/A\\
$\eta_c$ decays & N/A & N/A & 0.18 & $<0.01$ \\
$\eta_c$ width & N/A & N/A & 0.16 & $<0.01$ \\
$\eta_c$ line shape & N/A & N/A & N/A & 0.09 \\
\hline
Sum in quadrature & $\pm$0.15 & $\pm$0.14 & $\pm$0.52 & $\pm$0.77\\
\hline \hline
\end{tabular}
\end{center}
\end{table}

Systematic errors in the two analyses due to various possible sources were
estimated by varying the parameters used. These include choice of 
background parameterization, $\Gamma(h_c)=0.5-1.5$ MeV, 
$\pi^0$ line shape (varied by $\pm$10\%), 
bin size (varied between 0.5 and 2 MeV), 
$\pi^{0}$ energy calibration (varied energy of photon daughters 
by 0.2\% to 1.0\% depending on photon energy).
In the branching ratio for $\psi(2S) \to \gamma \eta_c$ \cite{hadron2007}
the dominant systematic uncertainty 
is due to the line shape of the $\eta_c$ which 
propagates into our product branching fraction analysis.
An  additional 2\% systematic uncertainty 
is included to account for the possibility that line
shape for the E1 transition $h_c \to \gamma\eta_c$ differs
from that for the M1 transition in direct $\psi(2S) \to \gamma \eta_c$
in a way that does not cancel in Eq. (2). 
It was determined that the results were stable 
well within statistical errors for the variations of event selection criteria.

The individual contributions to systematic errors, as well as their sum
in quadrature, are listed  in Table II.

When the exclusive events are removed from the inclusive spectrum, and the data are refitted, we obtain $M(h_c)=3525.35\pm0.27$(stat).  The average of this result for the (inclusive--exclusive) events and the result in Table 2 for the exclusive events gives our final result as
\begin{eqnarray}
M(h_c)= 3525.28\pm0.19(\mathrm{stat})\pm0.12(\mathrm{syst})\;\mathrm{MeV,}~~~~~~~~~\\
\nonumber\mathcal{B}_{1}(\psi(2S) \to \pi^{0} h_c) \times \mathcal{B}_{2}(h_c \to \gamma\
\eta_c)~~~~~~~~~~~~~~~~~~~~~~~~ \\  =(4.19\pm0.32\pm0.45)\times 10^{-4}.~~~~~~~~~~~~~~~~~~~~~~~~~~~~~~~
\end{eqnarray}
These results represent a large improvement over our earlier results.
The significance of $h_c$ identification is
 10 $\sigma$ for the inclusive measurements, and  
 13 $\sigma$ for the exclusive measurements.
The present results from the exclusive measurements are based on twice 
as many decay
channels of $\eta_c$ as before, and are in excellent agreement with
the results from the inclusive measurements. 

The nearly one order of magnitude larger statistics available in the
present measurements has enabled us to determine the systematic errors
presented in Table~II with much greater precision than in our earlier
publication \cite{hcold}. This allows us to average the present results with the
previous ones. The resulting average results are:
\begin{eqnarray}
      M(h_c)_{\mathrm{AVG}} = 3525.20\pm0.18\pm0.12~\mathrm{MeV},\\
(\mathcal{B}_1\times\mathcal{B}_2)_{\mathrm{AVG}} = (4.16\pm0.30\pm0.37)\times10^{-4}.
\end{eqnarray}

To put our results in perspective, we wish to make two further observations.

It is expected that the E1 radiative transitions 
$\chi_{c1} \to \gamma J/\psi$ and $h_c \to \gamma\eta_c$ should be similar.  Also, the total widths of $\Gamma(\chi_{c1})$ and  $\Gamma(h_c)$ should be similar.
If we assume them to be identical, it follows that
$\mathcal{B}_{2}(h_c \to \gamma\eta_c)=\mathcal{B}(\chi_{c1} \to \gamma J/\psi)\approx0.36\pm0.02$  \cite{pdg}. 
Our product branching fraction then leads to 
$\mathcal{B}_{1}(\psi(2S) \to \pi^{0} h_c)\approx(1.13\pm0.15)\times 10^{-3}$.
Incidentally, this is nearly equal to that for the only other 
isospin forbidden decay measured within the charmonium family,
$\mathcal{B}(\psi(2S) \to \pi^{0} J/\psi)$=(1.26$\pm$0.13)$\times 10^{-3}$.
A recent theoretical prediction \cite{kuang} gives the range
$\mathcal{B}(\psi(2S) \to \pi^{0} h_c)$=(0.4--1.3)$\times 10^{-3}$.

If the mass of the centroid of $^3P_J$ states, $\left<M(^3P_J)\right>$ is used as a measure of $M(^3P)$, the present measurement of $M(h_c)$ in Eq.~4 leads to
\begin{eqnarray}
\nonumber\Delta M_{hf}(1P)\equiv\left<M(^3P_J)\right> - M(^1P_1)~~~~~~~~~~~~~~~~~~~~~~~~\\
=+0.02\pm0.19(\mathrm{stat})\pm0.13(\mathrm{syst})\;\mathrm{MeV.}~~~~~~~~~~~~~~~~~~
\end{eqnarray}
The CLEO average mass in Eq.~(6) leads to
\begin{equation}
\Delta M_{hf}(1P)=+0.08\pm0.18(\mathrm{stat})\pm0.12(\mathrm{syst})\;\mathrm{MeV.}
\end{equation}
These results are consistent with the lowest order expectation of $1P$ hyperfine 
splitting being zero.
We notice that the triplet mass used above was obtained as 
$\left< M(^3P_J)\right> = (M(^3P_0)+3M(^3P_1)+5M(^3P_2))/9$, which 
is the evaluation of $M(^{3}P)$  in the lowest order,
when the spin--orbit splitting 
is perturbatively small.  
It has been pointed out \cite{martinrichards} that with
 $(M(^3P_2)-M(^3P_0))\approx140$~MeV, 
the validity of the perturbative determination of $M(^3P)$ is questionable.  Indeed, the perturbative prediction that $M(^3P_1)-M(^3P_0)=\frac{5}{2}[M(^3P_2)-M(^3P_1)]=113.9\pm0.3$~MeV disagrees with the experimental result, $95.9\pm0.4$~MeV, by 18~MeV.  This necessarily implies that the true $M(^3P)$ is different from the centroid value $\left<M(^3P_J)\right>$.  Since $\Delta M_{hf}(1P)$ is expected to be small ($\sim$~few~MeV), if not identically zero, it is important that higher order effects should be taken into account in deducing $M(^3P)$ from the known masses of $^3P_J$ states \cite{martinrichards}, so that a true measure of $\Delta M_{hf}(1P)$ can be obtained.
Only then can the present measurement of $M(h_c)$ be used to distinguish between the different potential model calculations, whose predictions for $\Delta M_{hf}(1P)$ vary over a large range because of the different assumptions they make about relativistic effects, the Lorentz nature of the confinement potential, and smearing of the spin--spin contact potential \cite{godfreyrosner}.
Although the presently available lattice calculations do not
have the required precision \cite{lattice}, 
it may be expected that future unquenched 
lattice calculations will resolve these problems.

We gratefully acknowledge the effort of the CESR staff in providing us  
with excellent luminosity and running conditions. This work was  
supported by the A.P.~Sloan Foundation, the National Science  
Foundation, the U.S. Department of Energy, the Natural Sciences and  
Engineering Research Council of Canada, and the U.K. Science and  
Technology Facilities Council.

\end{document}